\documentclass[aps,prb,twocolumn,groupedaddress,showpacs]{revtex4}
\usepackage{graphicx}
\usepackage{natbib}
\begin{document}

\title{Charge-imbalance effects in intrinsic Josephson systems}

\author{S.~Rother, Y.~Koval, P.~M\"uller}
\affiliation{Physikalisches Institut III, Universit\"at
Erlangen-N\"urnberg, Erlangen D-91058, Germany}
\author{R.~Kleiner}
\affiliation{Physikalisches Institut, Universit\"at T\"ubingen, D-72076
T\"ubingen, Germany}
\author{D.~A.~Ryndyk, J.~Keller}
\affiliation{Institut f\"ur Theoretische Physik,
Universit\"at Regensburg, D-93040 Regensburg, Germany}
\author{C.~Helm}
\affiliation{Institut f\"ur Theoretische Physik,
ETH Z\"urich, CH-8093 Z\"urich, Switzerland} 

\date{\today}

\begin{abstract}
  
We report on two types of experiments with intrinsic Josephson systems made
from layered high-T$_c$ superconductors which show clear evidence of 
nonequilibirum effects: 1. In 2-point measurements of IV-curves in the 
presence of high-frequency
radiation a shift of the voltage of Shapiro steps from the  canonical value
$V_s=h f/(2e)$ has been observed. 2. In the IV-curves
of double-mesa structures an influence of the current through one mesa on the
voltage measured on the other mesa is detected. Both effects can
be explained by charge-imbalance on the
superconducting layers produced by the quasi-particle current, and can be
described successfully by a recently developed theory of nonequilibrium
effects in intrinsic Josephson systems.

\end{abstract}


\maketitle

\section{Introduction}

In the strongly anisotropic cuprate superconductors
Bi$_2$Sr$_2$CaCu$_2$O$_{8+\delta}$ (BSCCO) and
Tl$_2$Ba$_2$Ca$_2$Cu$_3$O$_{10+\delta}$ (TBCCO) the CuO$_2$ layers
together with the intermediate material form a stack of Josephson
junctions. In the presence of a bias current perpendicular to the 
layers each junction of
the stack is either in the resistive or in
the superconducting state leading to the well-known multibranch 
structure of the  IV-curves. \cite{Kleiner92,Kleiner94,Yurgens96}

In the case of weakly coupled layers we expect that the bias current
generates charge accumulation on the layers between a resistive and
superconducting junction. As the current through a resistive junction is
carried mostly by quasi-particles, while the current through a barrrier in
the superconducting state is carried by Cooper-pairs, we expect two types of
non-equilibrium effects: 1. charge fluctuations of the superconducting
condensate, which can be expressed by a shift of the chemical potential of 
the condensate  and 2. charge-imbalance between electron- and hole-like  
quasiparticles.

Non-equilibrium effects in layered superconductors have been discussed in a
number of papers in various contexts with different methods and
approximations. \cite{Koyama96,Bulaevskii96a,Artemenko97,Ryndyk98,Preis98,
Shafranjuk99,Ryndyk99,Helm01} In a recent theoretical paper \cite{Ryndyk02} 
we have investigated in particular the consequences of 
non-equilibrium effects on
experiments with stationary currents. We found that in this case only the
charge-imbalance is important and leads to a change in the voltage, while the
shift of the chemical potential has no influence on the measured voltage. The
latter determines e.g. the dispersion of longitudinal Josephson plasma 
waves \cite{Koyama96}  and can be observed in some  optical   properties.
\cite{Helm02a,Helm02}   

In this  paper we report on new experiments which show clear evidence of
non-equilibrium effects for stationary currents and which can be explained by
charge-imbalance on the superconducting layers. 
In the first type of experiments Shapiro steps 
produced by high-frequency irradiation are measured in mesa structures of
BSCCO with gold contacts. Here
a shift $\delta V$ of the step-voltage $\Delta V_S= h f/(2e) -
\delta V$ from its canonical value $hf/(2e)$ is observed, 
which can be traced back to a change of the contact resistance due to
charge-imbalance on the first superconducting layer.
In another type of experiments current-voltage curves
are measured for two mesas structured close to each other on the same base
crystal. Here an influence of the current
through one mesa on the voltage  drop on the other mesa has been measured 
which can be explained by charge-imbalance on the first common
superconducting layer of the base crystal.
Both experiments allow to measure the charge-imbalance relaxation rate.

We start with a brief summary of the theory
\cite{Ryndyk02} which will be used in the following. Then the sample
preparation and the different experiments are described and discussed. 
From the experiments the charge-imbalance relaxation time will be
determined.

\section{Outline of the theory}

Let us consider a stack of superconducting layers $n=1,2 \dots$ with a
normal electrode $n=0$ on top. The basic quantity which determines the
Josephson effect of a junction between layer $n$ and $n+1$ is the  gauge
invariant phase difference \begin{equation} 
\gamma_{n,n+1}(t)= \chi_n(t) -
\chi_{n+1}(t) - \frac{2e}{\hbar} \int_n^{n+1} dz A_z(z,t) , 
\end{equation}
where $\chi_n(t)$ is the phase of the order parameter on layer $n$
and $A_z(z,t)$ is the vector potential in the barrier. 
 For the time
derivative of $\gamma_{n,n+1}$ one  obtains the general  Josephson relation:
\begin{equation} 
\dot \gamma_{n,n+1} = \frac{2e}{\hbar} \Bigl( V_{n,n+1} +
\Phi_{n+1} - \Phi_n\Bigr).  \label{genJos1}
\end{equation} 
Here  
\begin{equation}
V_{n,n+1}= \int_n^{n+1}dz E_z(z,t),
\end{equation} 
\begin{equation}
\Phi_n(t)= \phi_n(t) - \frac{\hbar}{2e} \dot \chi_n(t),
\end{equation}
are the voltage  and the gauge invariant scalar
potential, $\phi_n(t)$ is the electrical scalar potential. 
The quantity $e\Phi_n$
can be considered as shift of the chemical potential of the superconducting
condensate (in this paper the charge of the electron is written as $-e$).

The total charge fluctuation $\delta \rho_n$ on layer $n$ consists of charge
fluctuations of the condensate and of charge fluctuations of
quasi-particles. It is convenient to express the latter also by a kind of
potential $\Psi_n$ writing
\begin{equation}
\delta \rho_n = -2e^2N(0)(\Phi_n -\Psi_n) . \label{rho}
\end{equation}
With help of the Maxwell equation  ($d$ is
the distance between the layers, $\epsilon$ the dielectric constant of the
junction)  
\begin{equation}
\delta \rho_n= \frac{\epsilon \epsilon_0}{d} (V_{n,n+1}-V_{n-1,n})
\label{Maxwell} \end{equation}
the generalized Josephson relation now  reads:
\begin{eqnarray}
\frac{\hbar}{2e }\dot \gamma_{n,n+1} &=& (1+2 \alpha)
V_{n,n+1} - \alpha (V_{n-1,n} + V_{n+1,n+2}) \nonumber \\
&+& \Psi_{n+1} - \Psi_n
\label{genJos}
\end{eqnarray}
with $\alpha= \epsilon \epsilon_0/(2e^2N(0)d)$. It shows that the Josephson
oscillation frequency is determined not only by the voltage  in the same
junction but also by the voltages  in neighboring junctions. Furthermore it 
is influenced by the quasi-particle potential $\Psi$ on the
layers. If we neglect the latter we obtain for $\dot \gamma_{n,n+1}$ the same
result as in Ref.~\onlinecite{Koyama96}.

These equations for the voltage between the layers have to be  supplemented 
by an equation for the current density: in the stationary state (no
displacement current) it can be written as \cite{Ryndyk02} 
\begin{eqnarray} 
j_{n,n+1} = j_c \sin\gamma_{n,n+1} \nonumber \\
+\frac{\sigma_{n,n+1}}{d} \Bigl((1+2\alpha) V_{n,n+1} -\alpha(V_{n-1,n} +
V_{n+1,n+2}) \Bigr) . \label{current2}  
\end{eqnarray} 
Here the quasi-particle current between layers $n$ and $n+1$ is driven not 
only
by the voltage $V_{n,n+1}$ between the layers but also by additonal terms
which result from the charge fluctuation on the
two layers.  In the stationary state the dc-density $j_{n,n+1}$ is the
same for all barriers and is equal to the bias current density $j$. 
If the Josephson junction is in the
resistive state we may neglect the dc-component of the supercurrent
density in (\ref{current2}) for junctions with a
large McCumber parameter $\beta_c\gg1$. Then we 
obtain:
\begin{equation} 
\frac{j d}{\sigma_{n,n+1}} = 
(1+2\alpha) V_{n,n+1} -\alpha(V_{n-1,n} +
V_{n+1,n+2}). \label{current3}   
\end{equation} 

For the junction between the normal electrode ($n=0$) and the first 
superconducting layer ($n=1)$ we may neglect charge
fluctuations on the former and obtain for the current: 
\begin{equation}  
\frac{j d}{\sigma_{0,1}}=
(1+\alpha)V_{0,1} - \alpha V_{1,2}  . \label{contact}
\end{equation}  

Finally we need an equation of motion for the quasi-particle
charge. Here we consider a relaxation process between quasi-particle charge
and condensate charge within the layer. \cite{Clarke72, Schmid75, Tinkham96} 
In the stationary case the
quasi-particle charge is proportional to the charge-imbalance relaxation time
$\tau_q$ and the difference between supercurrents flowing in and out the
layer,  or equivalently, by the difference in quasi-particle currents:
\cite{Ryndyk02}
\begin{eqnarray}
%\Psi_n=(j_c \sin\gamma_{n,n+1} - j_c
%\sin\gamma_{n-1,n})d/\sigma_q = (j^{qp}_{n-1,n} - j^{qp}_{n,n+1})d/\sigma_q
\Psi_n &=&(j_c \sin\gamma_{n,n+1} - j_c
\sin\gamma_{n-1,n})\tau_q/(2e^2N(0)) \nonumber \\
&=& (j^{qp}_{n-1,n} - j^{qp}_{n,n+1})
\tau_q/(2e^2N(0)) \label{qprelax2} 
\end{eqnarray}
In the limit of small non-equilibrium
effects and for $T \ll T_c$ we may use the approximation $j^{qp}_{n,n+1}\simeq 
\sigma_{n,n+1}V_{n,n+1} = j$ for a resistive junction and $j^{qp}\simeq 0$
for a junction in the superconducting state. For example, if a current $j$
is flowing from a junction in the resistive
state into  a junction in the superconducting state, the charge-imbalance
potential generated on the superconducting layer between the two junctions
is $\Psi_n = j \tau_q/(2e^2N(0))$.

\section{Sample preparation and experimental set-up}

For measuring the intrinsic Josephson effect in BSCCO we used a
mesa geometry. Our base materials were BSCCO single crystals which
were grown by standard melting techniques. After glueing these
crystals on saphire substrates, a 100 nm thin gold layer was
thermally evaporated. The mesa was patterned by electron beam
lithography and etching by a neutral Ar atomic beam. As etching
rates for gold and BSCCO are known quite exactly, the number of
Josephson junctions inside the mesa could be adjusted to be
between 5 and 10.  To avoid shortcuts between Au leads and
the superconducting base crystal an insulating SiO layer was
evaporated. This material was removed from the top of the mesa by
liftoff. For contact leads a second Au layer was evaporated. The
shape of these leads was defined by standard photolithography and
etched by Ar atoms. For samples used in FIR experiments these Au
leads were formed in the shape of bow-tie antennas. \cite{Rother00} 

\begin{figure}
\includegraphics[width=14pc]{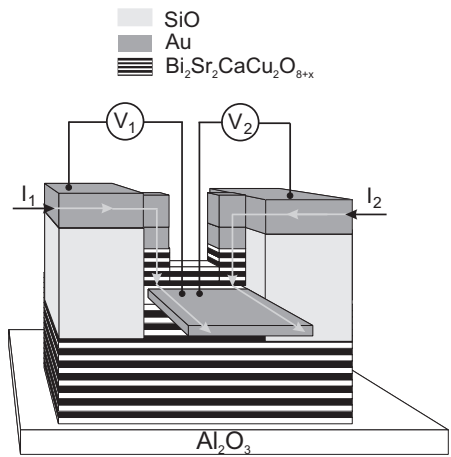}
\caption{Sketch of the sample geometry used for
injection experiments.
\label{fig:dmesa}}
\end{figure}

A second type of experiments is concerned with induced charge
imbalance. Here additional preparation steps were necessary.
These samples consist of two small mesas on top of one bigger
mesa (Fig.~\ref{fig:dmesa}).
%\begin{figure}
%\includegraphics[width=14pc]{dmesa}
%\caption{\label{fig:dmesa} Sketch of the sample geometry used for
%measurements described in ?.B.}
%\end{figure}
To realize this geometry the top layers of a larger mesa
(structured as described above) had to be cut in two smaller ones.
The whole sample was protected by electron-beam resist except a
thin line on top of the big mesa. By etching this sample the
larger mesa could be divided in two smaller mesas. To keep a part
of the base below the two smaller ones the etching process had to
be controlled precisely. The resistance of the gold lead was
recorded during etching showing the time when gold was removed and
separation of the big mesa began. With this technique we were able to
structure samples with $5 \times 10 \mu m^2$ and $4 \times 10 \mu
m^2$ mesas each including about 10 intrinsic Josephson junctions
on top of one larger one ($10 \times 10 \mu m^2$, 6 junctions). The gap
between the two top mesas was 1$\mu$m.  
All samples discussed in this article are listed in
Tab.~\ref{tab:table1}. The kind of experiments they are used for
are marked by DM (double mesa) and FIR
(FIR absorption).

\begin{table}

%\begin{ruledtabular}
\begin{tabular}{llcc}
Samples & area ($\mu m^2$)& $\#jj$ & experiments\\ \hline\hline
%SR103\_1& $5 \times 5$ & 11 & IV \\ ag2$\times$2\_jj11 & $2 \times
%2$ & 15 & IV\\ \hline 
SR102\_3 & $10 \times 10$ (B) & 6 & DM\\ &
$4 \times 10$ (M1) & 10 & DM\\ & $5 \times 10$ (M2)& 10 &
DM\\SH104 & $7 \times 6$ (B) & 11 & DM\\ & $2 \times 6$ (M1) & 3 &
DM\\ & $3 \times 4$ (M2) & 4 & DM\\ \hline \#32 & $8 \times 8$ & 8
& FIR\\ \#39 & $10 \times 10$ & 20 & FIR\\ \#20 & $10 \times 10$ &
11 & FIR\\
\end{tabular}
%\end{ruledtabular}
\caption{\label{tab:table1}Table of samples described in this
article. The area of the mesas and the number of junctions (\#jj)
containing it are listed  together with the kind of experiments they
were used for.}
\end{table}

IV-characteristics of our samples were recorded by applying dc
currents and recording voltages across mesas by digital
voltmeters. If not mentioned otherwise all measurements were
performed at a temperature of 4.2 K.

As radiation source for
experiments presented in Section 4.1 we used a far-infrared laser which
was optically pumped by a CO$_2$ laser. To minimize power losses
we used a polyethylene lens producing a parallel beam. Inside the
optical cryostat the samples were fixed in the center of a silicon
hyperhemispherical lens to focus the radiation onto the mesa.
\cite{Rother00} For some samples we used a very sensitive setup to
detect Shapiro steps and to determine voltage of these resonances
very precisely. For this purpose the laser beam was modulated by
an optical chopper with a fixed frequency. With the chopper
frequency as external reference, the voltage across the junction
was connected to the input of a lock-in amplifier (LI). As the LI
analyzes voltage changes occurring with the chopper frequency the
output signal $V_{LI}$ exhibits a point-symmetric structure. Thus
the LI output signal represents the voltage difference between IV-
characteristics with and without laser radiation.
 The voltage of the Shapiro step can
easily be identified as the symmetry point of this signal.

\section{Experimental results}
\subsection{Shapiro steps}

\begin{figure}
\includegraphics[width=20pc]{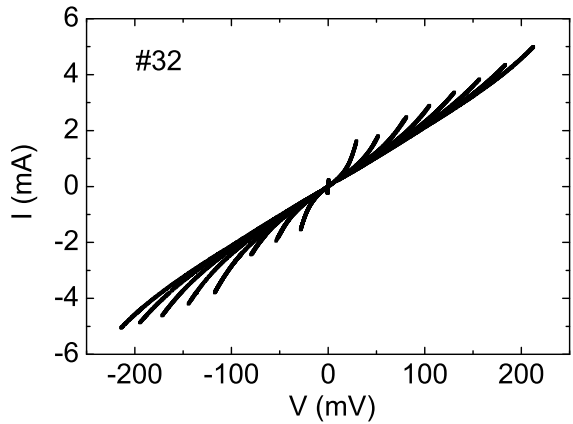}
\caption{\label{fig:32IV} IV characteristic of sample \#32.}
\end{figure}

Our experiments with high frequency electromagnetic radiation have been
carried out with three samples \#32 ($8\times 8 \mu m^2$),
\#39 ($10\times 10 \mu m^2$) and \#20 ($10\times 10 \mu m^2$). We first
discuss results for sample \#32 which consists of eight intrinsic Josephson
junctions as can be seen in the IV-characteristics shown in
Fig.~\ref{fig:32IV}.

The increasing values of critical currents may be explained by 
the inhomogeneous
etching process during fabrication producing layers with increasing areas from
top to bottom of the mesa. Then  it is natural to assume that the values of
critical currents increase with the position of the junctions inside the mesa. 
In particular, the first resistive branch of the IV-characteristics should be
assigned to the uppermost Josephson junction of the mesa. The critical current
on branch number 0 is strongly suppressed and its IV-curve is linear, while
the other branches show the typical non-linear IV-dependence characteristic
for a tunneling junction between two superconducting layers with d-wave order
parameter.  The special behaviour of branch 0 might be explained by the
assumption that the first superconducting layer is in proximity contact with 
the normal gold electrode.

\begin{figure}
\includegraphics[width=15pc]{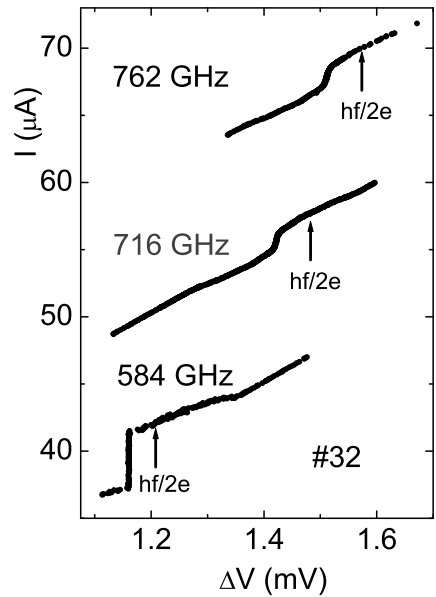}
\caption{\label{fig:32Vf} Voltages of Shapiro steps measured on IV
characteristic of sample \#32 compared to $hf/(2e)$ (for the higher
frequencies the current axis is shifted).}
\end{figure}

This sample was irradiated with
four external frequencies between 584 GHz and 762 GHz. On the IV
characteristic we could detect first order Shapiro steps on the 
first resistive branch at an absolute voltage $V_S$ and current $I_S$. To
compare the step-voltage with $hf/(2e)$ as predicted by the second
Josephson relation the contact voltage $V_0(I)$ between Au leads and
mesa in the superconducting state  without radiation had
to be measured. The value $\Delta
V_S=V_S(I_S)-V_0(I_S)$ can then be compared with $hf/(2e)$. As shown in 
Fig.~\ref{fig:32Vf}, $\Delta V_S$ evaluated for this sample is strictly lower
than $hf/(2e)$ for all four frequencies. The relative shift is
approximately $-3\%$.

This downshift can be explained if we assume that the Josephson
junction in the resistive state  which is locked to the external
radiation with frequency $f$ is close to the normal electrode, i.e.
between the  first and second superconducting layer. 
For this junction we have  
\begin{equation} \frac{hf}{2e}= (1+2\alpha)V_{1,2} -
\alpha(V_{0,1}+V_{2,3}) + \Psi_2-\Psi_1 , 
\end{equation}  
while for the other junctions, which are in the superconducting
state, we use 
\begin{equation}
0= (1+2\alpha)V_{n,n+1} - \alpha(V_{n-1,n}+V_{n+1,n+2}) +
\Psi_{n+1}-\Psi_n  \label{super}
\end{equation}  
for $n\ge 2$. Adding up these equations together with the current relation
for the contact with the normal electrode (\ref{contact}) 
we obtain for the total voltage:
\begin{equation}
V_S= \frac{jd}{\sigma_{0,1}}  + \frac{hf}{2e}.
\end{equation}
Here we have assumed that $\Psi_n=0$ for $n\ge 3$. The contribution of $\Psi_2$
drops out. Finally, the charge-imbalance potential $\Psi_1$ vanishes on
the first layer since the on- and off-flowing quasiparticle currents are
equal, $j^{qp}_{0,1}=j^{qp}_{1,2}=j$. 

We have to compare this result with the voltage measured in the absence of
high-frequency irradiation, when all junctions are in the superconducting
state. Then (\ref{super}) holds for $n\ge 1$. Adding up now equations
(\ref{super}) and (\ref{contact}) we obtain for the total voltage (contact
voltage in the superconducting state): 
\begin{equation}
V_0= \frac{jd}{\sigma_{0,1}} + \Psi_1.
\end{equation}
The quasi-particle potential on the first superconducting layer is now given
by $\Psi_1= j\tau_q/(2e^2N(0))$ while $\Psi_n=0$ for $n \ge 2$. 
Subtracting the measured contact voltage
from the voltage of the Shapiro step we obtain for the step-voltage:
\begin{equation}
\Delta V_S =  \frac{\hbar \omega}{2e} - \delta V  
\end{equation}   
with $\delta V = j\tau_q/(2e^2N(0))$. The shift is proportional to the 
life-time of charge-imbalance.  

\begin{figure}
\includegraphics[width=20pc]{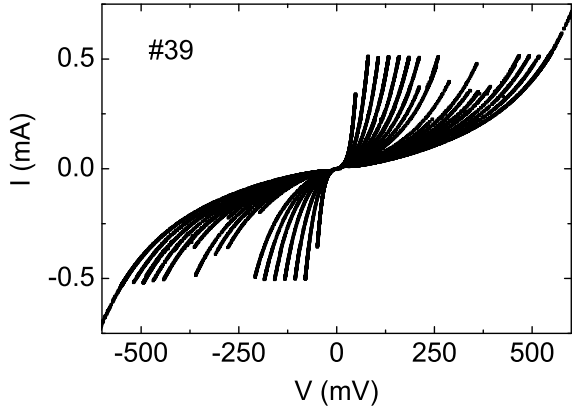}
\caption{\label{fig:39IV} IV characteristic of sample \#39.}
\end{figure}

Sample \#39 consists of 20 intrinsic Josephson junctions. For
this sample the distribution of critical currents is more homogeneous
(Fig.~\ref{fig:39IV}) than for \#32 making it impossible to
determine the position of the junction generating the first resistive
branch inside the mesa.
By using the sensitive measurement technique with a pulsed laser
beam we were able to detect Shapiro steps on the first resistive
branch at three FIR frequencies between 1.40 THz and 1.63 THz. The
voltage differences $\Delta V_S$ of these resonances were
calculated as described above. In contrast to \#32 for this sample
no deviations from $hf/(2e)$ were measured. In view   of
the  theory presented above this means, that the resistive junction which is
locked to external radiation is not close to the normal electrode or
the charge-imbalance time is rather short for this sample. 

\begin{figure}
\includegraphics[width=20pc]{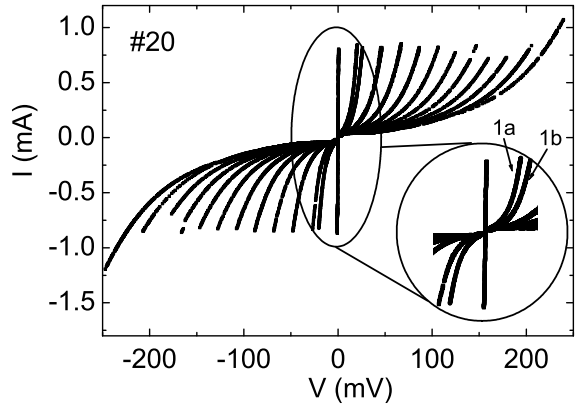}
\caption{\label{fig:20IV} IV characteristic of \#20. Several
current cycles are shown in one figure.}
\end{figure}

Finally we want to focus on sample \#20 consisting of 11 intrinsic
Josephson junctions. The IV-curves  of this sample show a very
homogeneous distribution of critical currents making it impossible
to assign any branch to a junction at a certain position
inside the mesa. However, for different current cycles (increase $I_1$ to its 
maximum value than decrease it zero) two  different first branches
marked 1a and 1b in Fig. \ref{fig:20IV} are measured. 
This means that two different junctions  in the stack become
resistive first.

\begin{figure}
\includegraphics[width=13pc]{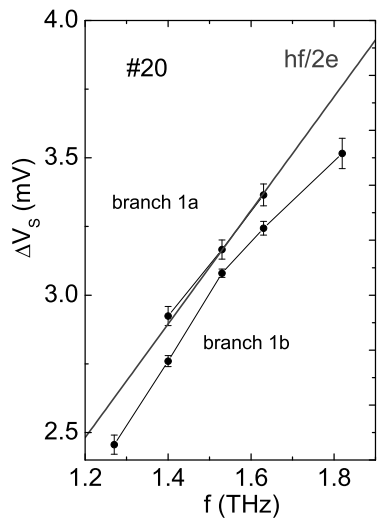}
\caption{\label{fig:20Vf} Voltages of Shapiro steps measured on IV
characteristic of sample \#20 compared to $hf/(2e)$. The values
strongly depend on which branch 1a or 1b the resonances occured.}
\end{figure}

We were able to detect Shapiro steps at five
frequencies between 1.27 THz and 1.82 THz on both branches 1a and 1b. Analyzing
the voltage differences $\Delta V_S$ we got different values for 1a and
1b (Fig.~\ref{fig:20Vf}). On branch 1a the voltages of Shapiro
steps were detected at regular values of $hf/(2e)$ whereas voltages of 1b were
shifted to values $3\%$ below $hf/(2e)$. This behaviour can be explained by
assuming that in case  1b the junction close to the normal electrode becomes
resistive while in case 1a a junction inside the stack becomes resistive. This
assumption is also in agreement with the observation that the voltage of the
second branch at the critical current is approximately given by the sum of the 
voltages of 1a and 1b
and the voltage differences for the higher branches are rather homogenous
and are close to the value for branch 1a.  


We want to mention that we could also measure Shapiro steps on
higher resistive branches. Due to slightly varying parameters of
different resistive Josephson junctions analysis of $\Delta
V_{s,n}$ is much more complicated and was not accurate enough to deduce
deviations from the second Josephson relation.

Finally we want to point out that a shift in the Shapiro
step voltage is only possible, if the resistive junction is close to the
normal electrode. In recent measurements of Shapiro steps in
step-edge junctions \cite{Wang01} this is different. Here the
resistive Josephson junction is inside a stack of superconducting
layers and hence no shift is observed. Also no shift of Shapiro steps appears 
in true 4-point measurements and for steps crossing the zero-current line at
the point of zero current. \cite{Doh00}

\subsection{Injection experiments in double-mesa structures}

The geometry of samples used for injection experiments in double-mesa
structures is schematically shown in Fig. \ref{fig:dmesa}. As results obtained 
for the different samples are rather similar, we will discuss only results for 
sample \#SR102\_3. Here 
two small mesas M1 of lateral size $5 \times 10 \mu m^2$ and M2
of size $4\times 10 \mu$m$^2$  were structured on a base mesa B of size $10m
\times 10 \mu m^2$. The currents $I_{1,2}$ through the mesas  M1 and M2
and the base mesa B and the corresponding voltages $V_{1,2}$ can be measured
separately. The brush-like structure of the IV-curves which is similar for
the two mesas shows two sets of branches with different critical currents.
These belong to 10 Josephson junctions in M1 and M2  and 6
junctions in the base mesa.

\begin{figure}
\includegraphics[width=20pc]{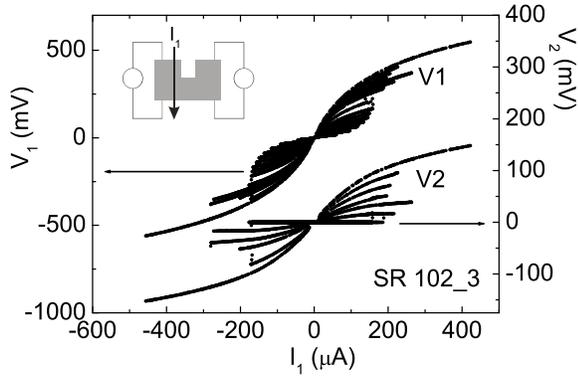}
\caption{\label{fig:1024point} Voltages of M1 and M2 during
variation of $I_1$ while $I_2=0$.}
\end{figure}

To explain the operation of the device we want to start with the case of no
current flowing through M2 ($I_2=0$). Measuring $V_1$ and $V_2$ during
variation of $I_1$ we obtained the curves shown in Fig.~\ref{fig:1024point}. 
$V_1(I_1)$ shows the full set of IV-curves of M1 and B, while in
$V_2(I_1)$ only the resistive junctions of the base mesa B appear. Here
$V_2(I_1)$ is in fact  a 4-point measurement of the IV-characteristic of the 
base mesa.
Note that as long as the junctions in B are completely superconducting, the
voltage $V_2$ is exactly zero and independent of $I_1$. If a
small current $I_2$ is applied to M2, then $V_2(I_2)$ shows the contact
resistance between the gold electrode and the mesa. Again the voltage $V_2$
does not depend on $I_1$.

\begin{figure}
\includegraphics[width=20pc]{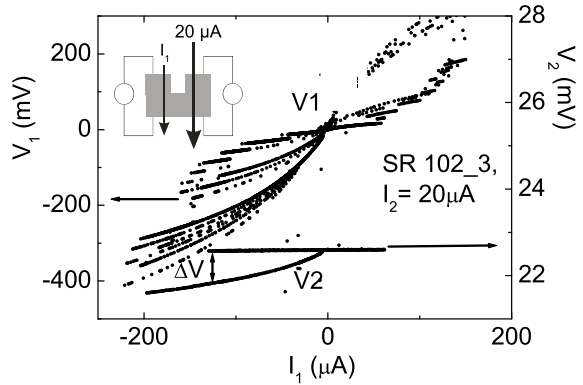}
\caption{\label{fig:102dV} Voltages of M1 and M2 during variation
of $I_1$ while two junctions of M2 are in the resistive state.}
\end{figure}

Now a larger bias current is applied to M2, such that some junctions of M2 are
in the resistive state. Keeping $I_2$ fixed we varied the current $I_1$.
During the cycling of $I_1$ junctions in M1 are switched on and
off into the resistive state. From time to time also junctions 
in the other mesa M2 are switched on and off leading to discrete jumps
in the voltage $V_2$, but otherwise $V_2$ is still constant (horizontal line
in Fig.~\ref{fig:102dV}).

In some cases, however, the voltage on M2 jumps to an additional branch
$V_2(I_1)$, which splits off from the constant voltage branch and depends
weakly  on $I_1$.  The jump into this branch, which is marked $\Delta V$ in
Fig.~\ref{fig:102dV}, is  always triggered by a switch into one of the higher
order branches of M1. Let us note that this happens  only if  $I_1$ and $I_2$
are in opposite direction.   In later cycles the branch $V_2(I_1)$ can also
traced out by increasing the current $I_1$ from zero. In general, several sets 
of voltage curves with split branches $\Delta V_2(I_1)$ occur. The voltages
of the horizontal branches correspond to different numbers of
junctions of M2 beeing in the resistive state at the fixed current $I_2$.
In Fig.~\ref{fig:102dV} only the branch corresponding to two resistive
junctions is shown.

%\begin{figure}
%\epsfig{figure=fig8.eps,width=25pc}
%\caption{\label{fig:102dV} Voltages of M1 and M2 during variation
%of $I_1$ while two junctions of M2 are in the resistive state.}
%\end{figure}

\begin{figure}
\includegraphics[width=16pc]{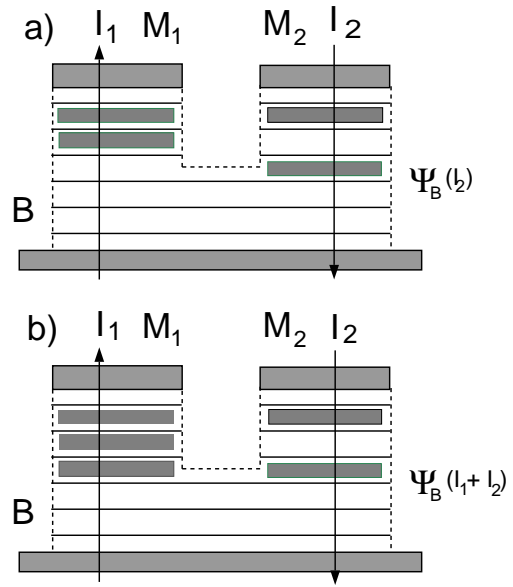}
\caption{\label{doublemesa} Charge-imbalance generated on the first
layer of the base mesa by the currents through mesas M1 and M2.} 
\end{figure}

The maximum voltage difference $\Delta V$ is much smaller than the
voltage between different resistive branches in M2 at the bias
current $I_2$. It is also much smaller than the voltage of resistive junctions
in the base mesa. Therefore the appearence of $\Delta V_2(I_1)$ must have a
different origin and can be explained as follows: If the junctions between the
lowest layers in both mesas M1 and M2 and the first common superconducting
layer in B are resistive, then a nonequilibrium potential $\Psi_B$ is
generated on the first superconducting layer of the base mesa. This 
is illustrated in Fig.~\ref{doublemesa}: In Fig.~\ref{doublemesa}a only the
current $I_2$ contributes to the charge-imbalance potential, while in
Fig.~\ref{doublemesa}b both currents contribute. The generated charge-imbalance
potential can be measured directly as additional  voltage on M2: 
\begin{equation} 
\Delta
V_2(I_1) =\Psi_B(I_1+I_2) - \Psi_B(I_2),
\end{equation}
\begin{equation}
\Psi_B(I)=
\frac{I}{A}\frac{\tau_q}{2e^2 N(0)} . \label{dmpot} 
\end{equation} 
Here $A$ is
the area of the base mesa.  In deriving (\ref{dmpot}) we assumed that the
charge-imbalance relaxation time is large compared to the diffusion time of
charge-imbalance along the layer.

Finally we want to explain the asymmetry of $V_2(I_1)$ with
respect to the polarity of both currents. The total current $I_B$
through B is either $I_1+I_2$ if the polarity is the same, or $I_1-I_2$ if
the polarity is opposite. In the first case the
total current through B is higher. This makes it
possible that one junction inside B gets resistive before the critical
current of the small mesas is exceeded. This destroys the precondition
of a completely superconducting mesa B. When the polarity is opposite 
the current through B is
always smaller than $I_1$ and junctions in M1 will get resistive before any
junction in B.

\subsection{Determination of the charge-imbalance relaxation time}

Both types of experiments can be used to measure the charge-imbalance
relaxation time.  From  the shift of the Shapiro step we obtain
\begin{equation}
\tau_q= \delta V \frac{A}{I} 2e^2 N(0) ,
\end{equation}
where $A$ is the area of the mesa. From the injection experiment in the
double-mesa structure a similar expression is obtained. 
This equation contains the unknown
density of states $N(0)$ of the two-dimensional electron gas at the Fermi
surface. For a rough estimate we may use the density of states $N(0)=
m/(2\pi\hbar^2)$ of free conduction electrons with mass $m$, then $2e^2N(0)=
0.67 $C$^2$J$^{-1} $m$^{-2}$. Alternatively, by using
$\alpha=\epsilon\epsilon_0/(2e^2N(0)d)$ we can express 
$2e^2N(0)$ by the  parameter $\alpha$, 
the dielectric constant $\epsilon$ of the barrier and the
distance $d$ between superconducting layers. In Ref.~\onlinecite{Helm02} the
values $\alpha=0.4$, $\epsilon=20$ have been estimated from reflectivity
experiments (for another material). Here it is found that the value of
$\epsilon \epsilon_0/(\alpha d)$ is rather close to the value of
$2e^2N(0)$ calculated for free conduction electrons, which therefore will be
used in the following.

With the data of sample \#32: $A= 64 \mu$m$^2$, $I= 30
\mu$A, $\delta V= 50 \mu$V we find $\tau_q \simeq 70$ ps. An
estimate for the shifts on branch 1b of  sample \#20 with 1.63 THz gives
values of $\tau_q \simeq$ 450 ps.
In a similar way we can also determine $\tau_q$ from the injection
experiments in double-mesa structures. Using $I_1=120\mu$A, $I_2 =
20 \mu$A, $\Delta V= 0.6$mV, $A= 100 \mu$m$^2$ we find $\tau_q
\simeq$ 330 ps, which is  similar to the values obtained from
the shifts of Shapiro steps. 

The charge-imbalance relaxation time $\tau_q$
describes the recombination of quasi-particles into Cooper pairs. It should be
distinguished from the characteristic time for the thermalisation of hot
quasi-particles, which is much shorter. \cite{Nessler98} Note that in the 
present
case $\tau_q$ is much longer than the period of a Josephson oscillation on a
Shapiro step.  For the charge-imbalance relaxation normally inelastic phonon
scattering processes are responsible. It should be noted that  in the case of
d-wave pairing with an anisotropic gap also elastic impurity scattering
contributes. \cite{Tinkham96}

\section{Summary}

In this paper we have discussed new experiments showing evidence of 
non-equilibrium effects in intrinsic Josephson
contacts in layered superconductors, which are due to charge-imbalance
produced by a stationary bias current. In particular, we
have investigated the voltage-position of Shapiro
steps in the presence of high-frequency irradiation. In some cases we
observed a down-shift of 3\% from the canonical value of $hf/(2e)$. This
shift, which is not a violation of the basic Josephson relation, can be
explained by charge imbalance on the first superconducting
layer if the Josephson contact is next to the normal electrode. 

In another type of experiment we studied the mutual influence of currents
through two mesas on a common base mesa on the measured voltages. The reults
can be explained by charge-imbalance on the first common superconducting layer
of the base mesa. This experiments has some similarity with the classical
experiment by Clarke, \cite{Clarke72} where charge imbalance is produced  in a
superconductor by a strong quasi-particle current which is then detected as
voltage difference between a normal contact and a Josephson contact. 

Both experiments allow to measure the charge-imbalance relaxation time which
is of the order of 100 ps. 
\section*{Acknowledgements:}

This work was supported by the Bayerische For\-schungs\-stiftung (S.R), the 
German
Science Foundation (D.R.), and by the Swiss National Center of Competence in
Research "Materials with Novel Electronic Properties-MaNEP" (C.H.).

\end{document}